\documentclass[aps,prl,twocolumn,superscriptaddress]{revtex4-2}

\usepackage{graphicx}
\usepackage{amssymb, amsmath}
\usepackage{xcolor}
\usepackage{multirow,tabularx}

\usepackage[utf8]{inputenc}

\newcommand{\be}{\begin{equation}}
\newcommand{\ee}{\end{equation}}

\newcommand{\ket}[1]{\left|#1\right\rangle}

\newcommand{\Xn}{\text{X}}
\newcommand{\Xp}{\text{X}^+}

\newcommand{\XXn}{\text{XX}}

\newcommand{\RepRate}{\Gamma_\text{rep}}
\newcommand{\TRep}{T_\text{rep}}
\newcommand{\PhotPairRate}{\Gamma_\text{pair}}
\newcommand{\PPhotPairRate}{\Gamma^\text{2P}_\text{pair}}
\newcommand{\PhotPairEff}{\eta^\text{1P}_\text{pair}}
\newcommand{\PPhotPairEff}{\eta^\text{2P}_\text{pair}}
\newcommand{\ExEff}{\eta_\text{ex}}
\newcommand{\DetEff}{\eta_\text{det}}
\newcommand{\FSS}{\Delta_\text{FSS}}
\newcommand{\Neg}{N}
\newcommand{\NegTD}{N^\text{(2P)}}
\newcommand{\NegFD}{N^\text{(4P)}}
\newcommand{\Fid}{f}
\newcommand{\FidTD}{f^\text{(2P)}}
\newcommand{\FidFD}{f^\text{(4P)}}
\newcommand{\XXtoX}{\XXn\mathord{-}\Xn}

\newcolumntype{Y}{>{\centering\arraybackslash}X}

\begin{document}

%Title of paper
\title{Maximally entangled and GHz-clocked on-demand photon pair source}

\author{Caspar Hopfmann}
\email[]{c.hopfmann@ifw-dresden.de}
%\homepage[www.ifw-dresden.de]{Your web page}
%\thanks{}
%\altaffiliation{}

\author{Weijie Nie}

\author{Nand Lal Sharma}

\author{Carmen Weigelt}
\affiliation{Institute for Integrative Nanosciences, Leibniz IFW Dresden, Helmholtzstraße 20, 01069 Dresden, Germany}

\author{Fei Ding}
\affiliation{Institut für Festkörperphysik, Leibniz Universität Hannover, Appelstraße 2, 30167 Hannover, Germany}

\author{Oliver G. Schmidt}
\affiliation{Institute for Integrative Nanosciences, Leibniz IFW Dresden, Helmholtzstraße 20, 01069 Dresden, Germany}

\affiliation{Material Systems for Nanoelectronics, Technische Universität Chemnitz, 09107 Chemnitz, Germany}

\affiliation{Nanophysics, Faculty of Physics and Würzburg-Dresden Cluster of Excellence ct.qmat, TU Dresden, 01062 Dresden, Germany}

%Collaboration name if desired (requires use of superscriptaddress
%option in \documentclass). \noaffiliation is required (may also be
%used with the \author command).
%\collaboration can be followed by \email, \homepage, \thanks as well.
%\collaboration{}
%\noaffiliation

\date{\today}

\begin{abstract}
	
	We present a $1$ GHz-clocked, maximally entangled and on-demand photon pair source based on droplet etched GaAs quantum dots using two-photon excitation. By employing these GaP microlens-enhanced devices in conjunction with their substantial brightness, raw entanglement fidelities of up to $0.95 \pm 0.01$ and post-selected photon indistinguishabilities of up to $0.93 \pm 0.01$, the suitability for quantum repeater based long range quantum entanglement distribution schemes is shown. Comprehensive investigations of a complete set of polarization selective two-photon correlations as well as time resolved Hong-Ou-Mandel interferences facilitate innovative methods that determine quantities such as photon extraction and excitation efficiencies as well as pure dephasing directly - opposed to commonly employed indirect techniques.	
	
\end{abstract}

% insert suggested keywords - APS authors don't need to do this
%\keywords{}

%\maketitle must follow title, authors, abstract, and keywords
\maketitle

% body of paper here - Use proper section commands
% References should be done using the \cite, \ref, and \label commands

\section*{Introduction}

On-demand sources of entangled photons are a fundamental enabling technology for future long range quantum network schemes \cite{Cirac1997, Kimble2008,  Simon2017}. While commonly employed sources of entangled photon pairs rely on intrinsically limited spontaneous down conversion processes \cite{Wang2016, Liao2018}, quantum emitter based sources promise greater potential - albeit with higher system complexity. Enormous progress towards practical implementations of quantum emitter based sources has been made in the past decade \cite{Ritter2012, Schwartz2016, Bhaskar2020}. However, applications of current entangled photon sources are still limited by the simultaneous realization of long quantum memory storage times as well as a set of critical parameters, which consist of the emitter clock rate $\RepRate$, entanglement fidelity $f$, source efficiency $\eta$, photon indistinguishability $I$ and single photon purity $g^{(2)}(0)$ \cite{Loock2019}. While each parameter has been addressed individually in specifically designed devices and experiments \cite{Gschrey2015, Jahn2015, Huber2017, Chen2018} demonstrating a source which combines all of these parameters simultaneously in one and the same device has proven illusive. For example, while atomic systems feature high entanglement fidelities and information storage times, repetition rates are fundamentally limited \cite{Koerber2018}. Nitrogen or Silicon vacancy centers in diamonds are exceptional quantum memories \cite{Abobeih2018}, but cannot easily serve as deterministic sources of entangled photon pairs. Semiconductor quantum dots (QDs) feature attractive properties such as interoperability with existing semiconductor miniaturization technology, flexibility in emission wavelength and high potential clock rates. Even though exploiting the latter in both theoretical entanglement distribution protocols \cite{Loock2019,Pseiner2020} and experimental implementations \cite{Shooter2020, Mueller2020} has been a strong focus of recent activities, a device that addresses all required parameters is sorely lacking. With the exception of the quantum memory storage, we aim to address last-mentioned point in this study. To this end we employ droplet etched GaAs QDs embedded in broadband optical antenna structure \cite{Chen2018} and driven by resonant pulsed two-photon excitation (TPE) \cite{Mueller2014} at $1$ GHz clock rates. Similar structures have recently been used to demonstrate wavelength tuneable entangled photon pair sources \cite{Chen2016} and single QD entanglement swapping \cite{Zopf2019, BassoBasset2019} compatible with atomic quantum memories based on Rb atoms \cite{Keil2017, Zopf2018}.

\section*{Results and Discussion}

\begin{figure}
	\includegraphics[width=1.0\linewidth]{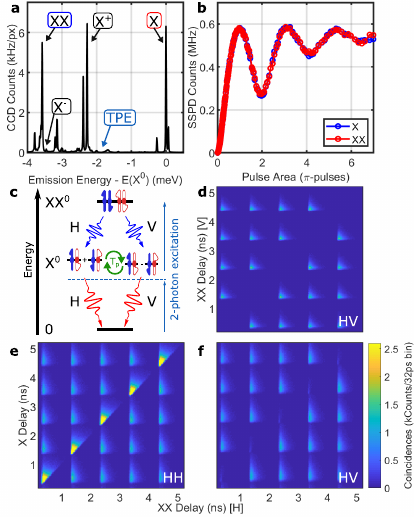}	
	\caption{$1.0$ GHz-clocked photon pair source using two-photon excitation (TPE). (a) High-resolution photoluminescence spectrum at $\pi$-pulse TPE with average power of $11$ $\mu$W. Basic QD transitions are annotated. (b) SSPD count rates as a function of excitation power. (c) Schematic illustration of the $\XXtoX$-cascade. Two-photon (d) auto- and (e,f) cross-correlation traces in detection polarization bases H and V.}
	\label{fig:basic_2PEx}	
\end{figure}

GHz-clocked photon pair emission is characterized by photoluminescence (PL), power dependent and single photon correlation spectroscopy in TPE, c.f. Fig. \ref{fig:basic_2PEx}(a), (b) and (d-f), respectively. Supplemented by polarization resolved (c.f. Suppl. Fig. S\ref{fig:Sup_FSS}) and power dependent PL spectroscopy, basic transitions of the QD emission spectra can be identified. Additionally, an excitonic fine structure splitting $\FSS = (3.9 \pm 0.3)$ $\mu$eV is observed. Even though one would expect only exciton ($\Xn$, $1.5910$ eV) and biexciton ($\XXn$, $1.5871$ eV) transitions to be observable in TPE, we find that this is not the case here. This can be attributed to the fact that the positive trion transition $\Xp$ is very close to excitation laser energy, which leads to parasitic excitation of positively charged QD states. The latter also causes an average positive charging of the QD, thereby reducing the TPE efficiency. Notably, this is the case even though the TPE power dependence of Fig. \ref{fig:basic_2PEx}(b) clearly demonstrates coherent driving of $\XXn$ through observation of Rabi oscillations \cite{Mueller2014}. $\pi$-pulses are achieved at pulse energies of $11.4 \pm 0.1$ fJ at which a photon pair rate $\PhotPairRate$ of $0.58$ MHz is recorded. Figs. \ref{fig:basic_2PEx}(d-f) depict $\XXn$ auto-correlation as well as linearly co- and cross-polarized cross-correlation $\XXtoX$ traces, respectively, recorded by time tagged (TTR) photon correlation. A clear anti-bunching is observed in $\XXn$ autocorrelation trace, from which a single photon purity of $(99.7 \pm 0.2)$ $\%$ and the $\XXn$ lifetime $T^{\XXn}_1 = (134.96 \pm 0.50)$ ps are deduced. The $\XXtoX$ co (cross)-polarized cross-correlation traces demonstrate a clear bunching (anti-bunching), as expected from the $\XXtoX$-cascade, c.f. Fig. \ref{fig:basic_2PEx}(c). The $\Xn$ lifetime $T^{\Xn}_1 = (200.4 \pm 1.4)$ ps remains far below the excitation cycle time of $\TRep \simeq 1.0$ ns, unambiguously demonstrating the on-demand behavior of the photon pair source at $\RepRate \simeq 1.0$ GHz.\\

While the total efficiency of the photon pair source can be determined straightforwardly by  $\PhotPairEff = \PhotPairRate / \RepRate = \ExEff \DetEff \simeq (8.1 \pm 0.4) \cdot 10^{-4}$, the actual values for the excitation $\ExEff$ and detection $\DetEff$ efficiencies cannot be easily determined without resorting to indirect methods such as simulations or calibration of optical losses. Another method that was proposed by \emph{Jahn et. al.} is to use the decay of the auto-correlation function to gain access to the QD transition excitation efficiency though characterization of its blinking behavior \cite{Jahn2015}. Such an analysis is performed and yields a maximum expected $\ExEff$ of $0.61 \pm 0.02$, c.f. Suppl. Fig. S\ref{fig:Sup_XX0_decay}. Here however a different method to determine $\ExEff$ and $\DetEff$ with no assumptions on the time dependence of the underlying QD carrier jitter processes or any other quantity in the experimental setup is proposed. The assumption is that for any one photon detection event of $\Xn$ or $\XXn$ a second photon is created for the same excitation pulse. This assumption is justified by the very high single photon purity as well the deterministic behavior of the $\XXtoX$ cascade in TPE. The co-polarized two-photon efficiency is therefore $\PPhotPairEff = \PPhotPairRate / \RepRate = \PhotPairEff \DetEff$, which facilitates interference of $\ExEff = (\PhotPairEff)^2/\PPhotPairEff = (7.6 \pm 0.9)$ $\%$ and $\DetEff = \PPhotPairEff/\PhotPairEff = (1.1 \pm 0.1)$ $\%$. The resulting $\ExEff$ is in strong contrast to the value deduced by the blinking analysis. Due to the simplicity of our model and the avoidance of any assumptions on the jitter, we are confident that the results describe the given photon pair source more accurately. The value determined for $\DetEff$ is in good agreement with our previous work, in which an equivalent photon extraction scheme was employed, the value of $\DetEff$ was however deduced indirectly from calibration of the experimental setup \cite{Zopf2019}.

\onecolumngrid

\begin{figure}
	\includegraphics[width=1.0\linewidth]{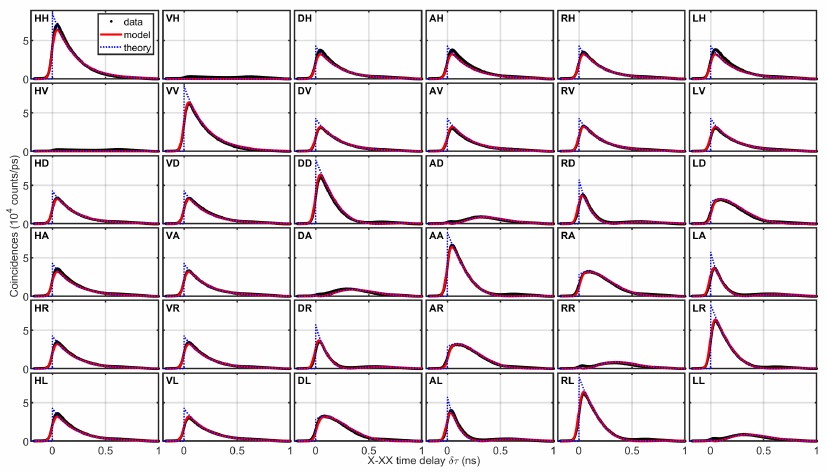}
	
	\caption{Matrix of polarization dependent two-photon correlation traces of detected time delay differences $\delta\tau = \tau^{I}_{\Xn} - \tau^{J}_{\XXn}$. Polarization combinations (IJ) of $\XXn$ and $\Xn$, respectively, are annotated. Experimental data ($1$ ps binning) taken with 4 detectors in TTR mode, fitted model and theoretical curves are shown. Details are given in the text. }
	\label{fig:corr_matrix}
\end{figure}

\twocolumngrid

In order to unambiguously demonstrate the maximally entangled nature of the QD-based entangled photon pair source two-photon correlation traces between all four detection channels and \emph{all} polarization base combinations are collected and modeled according to \emph{Winik et. al.} \cite{Winik2017}. Fig. \ref{fig:corr_matrix} depicts the resulting correlation traces, modeling as well as the underlying theoretical curve. The model of the precessing $\Xn$ state around its $H$ and $V$ polarization eigenbases fits the data exceptionally well, the mean square error is $1.5$ $\%$. All graphs of Fig. \ref{fig:corr_matrix} are modeled simultaneously and the only free parameter is a phase-offset $\delta \phi = (20 \pm 4)^{\circ}$. We attribute the observed $\delta \phi \neq 0$ to an alignment offset of the excitation polarization base with respect to the $\Xn$ eigenbases. Other model parameters, i.e.  $T_p = h/\FSS \simeq 1.05$ ns and $T^{\Xn}_1$, are determined independently.

\begin{figure}
	\includegraphics[width=1.0\linewidth]{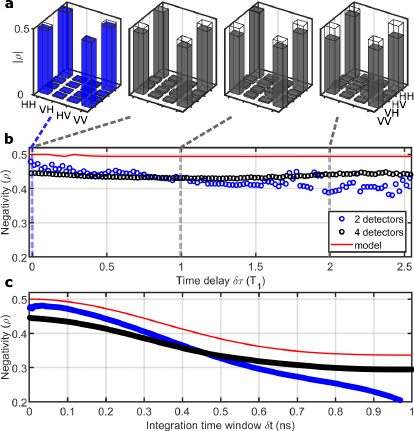}
	
	\caption{Two-photon entanglement extracted from polarization resolved correlation traces using 8 ps binning. (a) Exemplary representations of the two-photon density matrix elements $|\rho|$ for selected time delays and correlation modes. (b) Entanglement negativity $N(\rho)$ as a function of time delay difference $\delta\tau$ for fitted model as well as for two and four detector based correlation matrices utilizing histogram and TTR measurement modes, respectively. (c) $N(\rho)$ plotted against the correlation integration time window $\delta t$.}
	\label{fig:neg_delay}
\end{figure}

 From the polarization resolved correlation matrix the two-photon density matrix $\rho$ is calculated \cite{James2001}. Fig. \ref{fig:neg_delay}(a) illustrates $\rho$ as a function of selected $\XXtoX$ event time delays $\delta \tau$, while the corresponding entanglement negativities $\Neg(\rho)$ are shown in Fig. \ref{fig:neg_delay}(b). $\Neg(\rho, \delta \tau)$ is extracted from two-photon TTR-based four detector correlation data of Fig. \ref{fig:corr_matrix}, correlation histograms involving one detector pair (raw data shown in Suppl. Fig. S\ref{fig:P2_corr_matrix}) and the fitted model. Even though the fitted model uses the same limited detector timing resolution it exhibits maximally entangled behavior, independent of $\delta \tau$. The latter is also true for both two- and four-detector data, although with a lower absolute $\NegTD_\text{max} = 0.479 \pm 0.005$ ($\FidTD_\text{max} = 0.95 \pm 0.01$) and $\NegFD_\text{max} = 0.443 \pm 0.005$ ($\FidFD_\text{max} = 0.84 \pm 0.01$), respectively, where $\Fid_\text{max}$ denotes the corresponding entanglement fidelity. By leveraging the excellent agreement of the model and correlation data as well as the observed dependence on the selected polarization basis it may be inferred that $\Neg_\text{max}$ and $\Fid_\text{max}$ are limited only by the experimentally achievable purity of the polarization projection bases and not by the actual quantum state. This fact also explains the differences between the two and four detector data: When using two detectors one may choose a basis pair which yields the smallest projection errors, while this is not possible when all detector combinations are employed, thereby yielding increased projection errors. More information can be found in the supplementary material. Due to the four fold increased number of coincidences using four detectors in TTR mode compared to two detectors in histogram mode, the statistical errors are significantly reduced in the former case. The negativity remains practically constant for time delays up to $2.5 \, T^{\Xn}_1$ and far above the entanglement condition $N > 0$, which unambiguously demonstrates that the entanglement is not affected by the exciton fine structure splitting. For the two detector data $N(\delta \tau)$ drops slightly towards increasing $\delta \tau$, this may be attributed to the finite cross talk between adjacent pump periods predominately present in the  histogram mode correlation measurements (c.f. Suppl. Fig. S\ref{fig:P2_corr_matrix}). Fig. \ref{fig:neg_delay}(c) depicts $N(\rho)$ as a function of the time delay integration window  $\delta t$ for fitted model, two and four detector data. Due to $T_p >> T^{\Xn}_1$, $\NegFD(\delta t)$ does not fully oscillate, but levels off at about $\Neg(\delta t \to \TRep) \simeq 0.3$. $\NegTD(\delta t)$ does not level off, this again can be understood when considering the finite cross talk to adjacent pump periods in this measurement mode. $\Neg(\delta t)$ demonstrates why the precession of $\Xn$ needs to be taken into account by time resolved detection to achieve maximally entangled photon pair sources exhibiting $\FSS > 0$.

\begin{figure}[!h]
	\includegraphics[width=1.0\linewidth]{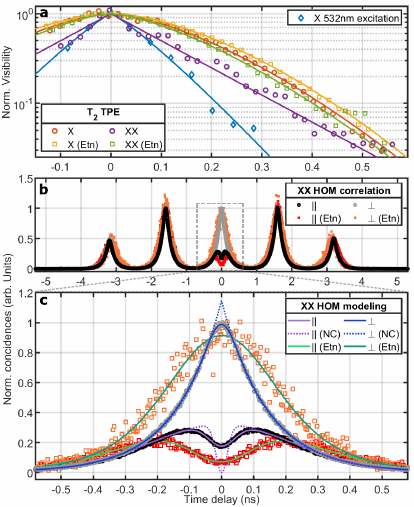}
	
	\caption{Photon coherence investigations using (a) Michelson (MI) and (b,c) Hong-Ou-Mandel (HOM) interferometers with and without employing an etalon (Etn) of $2$ GHz bandwidth. In the former the $T_2$ values for $\Xn$ and $\XXn$ are determined by modeling of the MI visibility of the respective experimental data to Lorentzian and Gaussian product line shapes. (b) Co- and cross-polarized ($\parallel$ and $\perp$) HOM two-photon interference traces. (c) Close-up of the central HOM interference manifold as well as the corresponding fitted model functions (see text) plotted against the two-photon interference time delay. The corresponding non-time-convoluted (NC) model functions are drawn as dotted lines.}
	\label{fig:coherence}
\end{figure}

In order to show the practical usefulness of the GHz entanglement sources in quantum interference schemes, such as entanglement swapping, high photon coherence and indistinguishability are indispensable. Measurements of $T_2$ coherence times and $\XXn$ indistinguishabilities employing Michelson and Hong-Ou-Mandel interferometers are depicted in Figs. \ref{fig:coherence}(a) and (b,c), respectively. To enhance the spectral resolution of these investigations $2$ GHz etalons are employed. As may be inferred from the Michelson visibly, TPE enhances the above band excitation coherence $T^{\Xn}_2(\text{532nm}) = (102 \pm 10)$ ps to $T^{\Xn}_2(\text{TPE}) = (242 \pm 7)$ ps and $T^{\Xn}_2(\text{TPE, Etn}) = (266 \pm 3)$ ps $\simeq 1.32 \, T^{\Xn}_1$ without and with etalon, respectively. While the usage of the etalon moderately enhances $T_2$, it effectively quenches beating of the visibility due to parasitic transitions (c.f. Fig. \ref{fig:basic_2PEx}(a)). For $\XXn$ $T^{\XXn}_2(\text{TPE}) = (144 \pm 24)$ ps and $T^{\XXn}_2(\text{TPE, Etn}) = (226 \pm 7)$ ps $\simeq 1.67 \, T^{\XXn}_1$ are obtained. Hong-Ou-Mandel interference traces recorded by $76$ MHz TPE with and without etalon as well as using co- and cross-polarized linear detection bases are shown in Figs. \ref{fig:coherence}(b) and (c), where the latter depicts a detailed view of the central interference manifold. Integrated and post-selected indistinguishabilities (using etalons) equate to $I_\text{itgr}^{\XXn} = 0.478 \pm 0.005$ ($I_\text{itgr}^{\XXn, \text{Etn}} = 0.67 \pm 0.02$) and $I_\text{ps}^{\XXn} = 0.829 \pm 0.001$ ($I_\text{ps}^{\XXn, \text{Etn}} = 0.93 \pm 0.01$), respectively, thereby demonstrating strong two-photon interference visibilities needed for practical applications.  Contrary to comparable indistinguishability studies \cite{Gschrey2015, Huber2017, Zopf2019} on QD systems the two-photon interference traces are modeled as a function of the time delay according to the theory provided by \emph{Legero et. al.} \cite{Legero2003}, with the pure dephasing time $T^*_2$ as the only free parameter. This model facilitates the direct extraction of $T^*_2$ - a quantity otherwise only indirectly accessible, i.e. by using $T_2 = (\frac{1}{2 T_1} + \frac{1}{\tilde{T^*_2}})^{-1}$. The modeling yields values with and without etalon of $T^*_2(\XXn, \text{Etn}) = (1054 \pm 38)$ ps and $T^*_2(\XXn) = (392 \pm 4)$ ps, respectively. When compared to the values of $\tilde{T^*_2}(\XXn, \text{Etn}) = (2040 \pm 590)$ ps and $\tilde{T^*_2}(\XXn) = (330 \pm 130)$ obtained indirectly, it becomes clear that even though the values match reasonably the direct method is by far more accurate and should therefore be used to characterize the pure dephasing.

\section*{Methods}

\begin{figure}
	\includegraphics[width=1.0\linewidth]{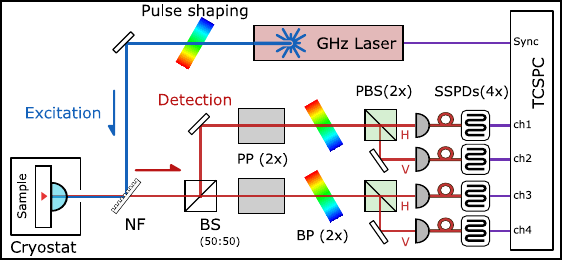}
	
	\caption{Schematic illustration of the experimental setup. The following abbreviations are used: Notch filter (NF), beam splitter (BS), polarization projection (PP), bandpass (BP), polarizing beam splitter (PBS), superconducting single photon nano wire detector (SSPD) and time correlated single photon counting (TCSPC).}
	\label{fig:Setup}
\end{figure}

As entangled photon pair sources droplet etched GaAs QDs embedded in a $Al_{0.15}Ga_{0.85}As$ matrix are employed \cite{Wang2007}. The GaAs/AlGaAs hetrostructure is grown on [001] GaAs substrate by molecular beam epitaxy. In order to enhance the photon extraction efficiency QD-nanomembranes are etched from the substrate, the resulting membranes are gold coated and attached to GaP microlenses in a flip-chip process \cite{Chen2018}. Consequently, the devices are glued to Si substrates and cooled to $3.8$ K using a exchange gas closed-cycle He refrigerator system. GHz TPE of the QDs is achieved using a tuneable $60$ fs-pulsed ($20$ nm bandwidth) laser system with a repetition rate of  $\RepRate = 0.9925$ GHz. The pulses are shaped by optical bandpasses and dispersion gratings to a bandwidth of $0.06$ nm ($5.4$ ps pulse length) and polarized using liquid crystal retarders and a linear polarizer, c.f. Fig. \ref{fig:Setup}. Laser excitation and QD photo luminescence is separated by volume Bragg grating notch filters. Emitted photons are collected by an aspheric lens as well as split into two independent detection paths. The polarization bases and bandpass filters of both arms may be controlled independently and are coupled each to two single mode fibers collecting orthogonal polarization bases. Time-resolved single photons are observed with superconducting nano wire detectors (SSPDs) and analyzed by employing correlation electronics (TCSPC) with combined timing resolutions of $47$, $56$, $54$ and $50$ ps for four detection channels, respectively. Time resolved photon correlation spectroscopy may be performed using direct two-photon correlation histograms on the TCSPC hardware or by using time tagged recordings (TTR) and software-based processing.   High resolution ($15$ $\mu$eV) spectroscopic studies are performed using a fiber coupled double spectrometer of $0.75$ m focal length and an attached charge coupled device (CCD). Polarization resolved detection is provided by liquid crystal retarders and polarizing beam splitters, they are calibrated to the QD polarization eigenbases with an accuracy of better than $0.95$. Hong-Ou-Mandel interference measurements are conducted by using the two fiber couplings of one detection arm and inserting a fiber based delay of $(1.58 \pm 0.02)$ ns followed by a fiber beam splitter. The excitation pulses are evenly split and delayed correspondingly. $T_2$ coherence times are determined using a fiber coupled Michelson interferometer.

\section*{Conclusions}

In conclusion this study demonstrates a GHz-clocked maximally entangled and on-demand photon pair source compatible to 780nm Rb-based atomic quantum memories. The raw entanglement negativities (fidelities) of up to $0.479 \pm 0.005$ ($0.95 \pm  0.01$) are not limited by either the $\Xn$ fine structure or the $\XXtoX$-cascade process but only by the experimental quality of the polarization projection of the two-photon correlations. Contrary to recent publications claiming GHz-clocked entangled photon pair sources \cite{Mueller2020, Shooter2020} no reset of the $\XXtoX$-cascade is required, which is also a necessary condition for fully on-demand entangled photon pair sources. We present a significantly enhanced method of determining $\ExEff$ and $\DetEff$ directly from two-photon correlation data with a minimal set of assumptions. This is especially true in regard to assumptions on setup efficiency and transition jitter time dependence employed by established methods \cite{Mueller2014, Jahn2015}. Contrary to what has been claimed concerning the latter, we show that the observation of Rabi oscillations as a function of excitation pulse energy is not a sufficient criterion for near unity excitation efficiencies. Time resolved investigations of Hong-Ou-Mandel two-photon correlations photon reveal indistinguishabilities of up to $0.93$ and $0.67$, with and without post selection, respectively. This paves the way for applications that rely on efficient two-photon interference - such as Bell state projection-based quantum repeaters schemes and photon graph state generation through photon fusion \cite{Wang2016}. Additionally, we employ for the first time a direct observation method of photon pure dephasing times $T_2^*$ yielding values up $5 \, T_1$ for the $\XXn$ transition by modeling the time dependence of the central Hong-Ou-Mandel interference manifold. The results represent an important step forward for GaAs QD-based entangled photon pair sources by drastically enhancing critical parameters such as clock rate and maximal entanglement irrespective of fine structure, all while maintaining acceptable photon indistinguishabilities and on-demand characteristics. This will lay the foundation to realize quantum repeater schemes exploiting the high QD clock rates as compared to other implementations of entangled photon pair sources \cite{Loock2019}. Future efforts with respect to long distance entanglement distribution will need to also take into account the challenge of quantum memories with long coherence times. Possible solutions could entail coupling entangled photon pair from QDs to atomic or diamond defect based memories \cite{Li2015} by using quantum frequency conversion \cite{Weber2019}, thereby combining the advantages of each material system in a mutually beneficial way. Another option could be coupling to long-lived nuclear spins \cite{Chekhovich2020}.

\begin{acknowledgments}
	We acknowledge funding by the BMBF (Q.link.X) and the European Research Council (QD-NOMS). We thank Michael Zopf and Jingzhong Yang (LU Hannover) as well as Robert Keil (IAF Freiburg) for fruitful discussions.	
\end{acknowledgments}

\section*{Supplementary Information}

\subsection*{Fine Structure Splitting}

\begin{figure}
	\includegraphics[width=1.0\linewidth]{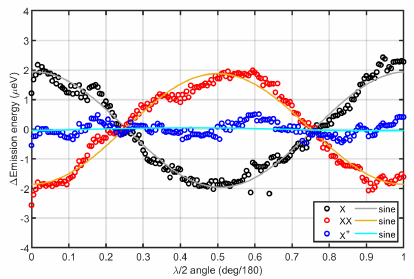}
	
	\caption{Relative emission energies of the $\Xn$, $\XXn$ and $\Xp$ transitions as a function of the linear polarization projection angle. The latter is achieved by rotation of a lambda half retarder plate combined with a fixed linear polarizer. The data is modeled by sine functions.}
	\label{fig:Sup_FSS}
\end{figure}

The $\Xn$ fine structure splitting ($\FSS$) of $\ket{H} = \frac{1}{\sqrt{2}} \ket{\uparrow \Downarrow} + \ket{\downarrow \Uparrow}$ and $\ket{V} = \frac{1}{\sqrt{2}} \ket{\uparrow \Downarrow} - \ket{\downarrow \Uparrow}$ is revealed by its characteristic linear polarization dependence on the central emission energy \cite{Bayer2002}. This can experimentally be realized by recording PL emission spectra as a function of the angle of a lambda half plate followed by a fixed polarizer. The emission energies of $\Xn$, $\XXn$ and $\Xp$ of such an experiment are depicted in Suppl. Fig. S\ref{fig:Sup_FSS} together with respective modeling to sine functions. From the latter $\FSS(\Xn) = (3.9 \pm 0.3)$ $\mu$eV, $\FSS(\XXn) = (3.7 \pm 0.3)$ $\mu$eV and $\FSS(\Xp) = (0.1 \pm 0.4)$ $\mu$eV are obtained.\\

\subsection*{Lifetime}

\begin{figure}
	\includegraphics[width=1.0\linewidth]{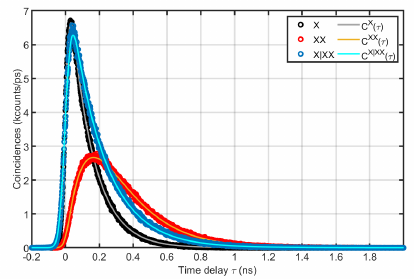}
	
	\caption{Time-resolved photon correlation traces ($1$ ps bin) extracted from two-correlation TTR data in H polarization base and $\RepRate = 76$ MHz. Modeling to exponential decays $C(\tau)$ according to equations given in the text.}
	\label{fig:Sup_Lifetime}
\end{figure}

The lifetime of QD transitions is typically inferred by recording the time-resolved photon emission correlation traces triggered by resonant excitation pulsed. We employ two-photon correlation data recorded in the H polarization bases using $76$ MHz TPE to determine the respective $\Xn$ and $\XXn$ lifetimes. The resulting correlation traces are depicted in Suppl. Fig. S\ref{fig:Sup_Lifetime}. Modeling correlation functions $C(\tau)$ are given by

\begin{align}
	C^{\XXn}(\tau) &\propto e^{-\tau/T_1^{\XXn}} \ast \mathcal{N}(\text{tres}_\text{ch1}) \, , \\
	C^{\Xn}(\tau)  &\propto e^{-\tau/T_1^{\XXn}} \ast e^{-\tau/T_1^{\Xn}} \ast \mathcal{N}(\text{tres}_\text{ch2}) \, , \\ 
	C^{\Xn|\XXn}(\tau)  &\propto e^{-\tau/T_1^{\Xn}} \ast \mathcal{N}(\sqrt{\text{tres}_\text{ch1}^2+\text{tres}_\text{ch2}^2}) \, .
\end{align}

Where $\mathcal{N}(\text{tres}_\text{ch})$ are normal distributions with a FWHM of $\text{tres}_\text{ch}$ representing the SSPD timing jitter determined independently. The only free parameters are the lifetimes, which equate to $T_1^{\XXn} = (135.0 \pm 0.5)$, $T_1^{\Xn} = (200 \pm 1)$ and $T_1^{\Xn|\XXn} = (197 \pm 1)$ ps. As expected from a near ideal $\XXtoX$-cascade $T_1^{\Xn} \simeq T_1^{\Xn|\XXn}$.\\

\subsection*{Two detector correlation matrix}

\onecolumngrid

\begin{figure}
	\includegraphics[width=1.0\linewidth]{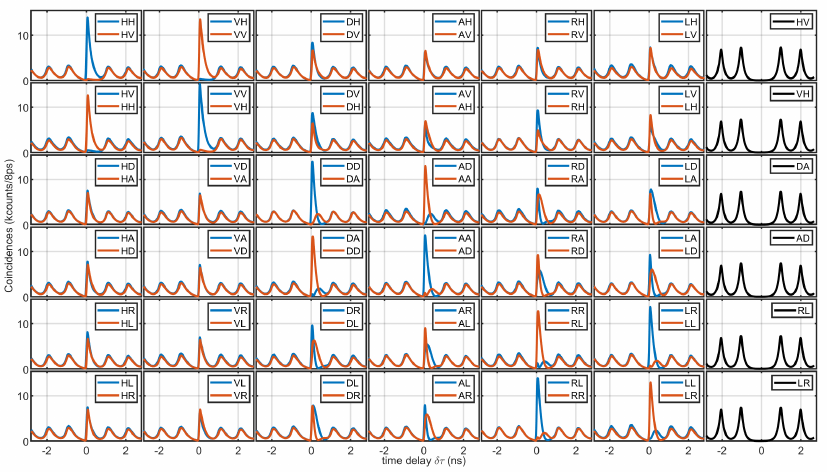}
	
	\caption{Matrix of polarization bases (IJ) dependent two-photon cross- and auto-correlation traces plotted as a function of photon arrival time delay differences $\delta\tau_\text{cross} = \tau^I_{\Xn} - \tau^J_{\XXn}$ and $\delta\tau_\text{auto} = \tau^I_{\XXn} - \tau^J_{\XXn}$, respectively. Data is recorded with different combinations of detectors in histogram mode. Cross-correlations between channel combinations 1 vs. 3 and 1 vs. 4 are drawn by red and blue lines, respectively, while auto-correlation traces (channel 1 vs. 2) are shown in black. The correlation traces of the latter are scaled by a factor of $\times 1/6$ as the polarization bases of channel 1 and 2 are necessarily complementary - i.e. the 36 polarization bases pairs of the cross-correlation matrices reduce to 6 distinct auto-correlation bases pairs.}
	\label{fig:P2_corr_matrix}
\end{figure}

\twocolumngrid

$\XXtoX$ two-photon correlation traces recorded by TSPC histograms are depicted in Suppl. Fig. S\ref{fig:P2_corr_matrix} for different detector combinations. Contrary to the four detector TTR correlation data, c.f. Fig. \ref{fig:corr_matrix}, there is a significant cross-talk in the cross-correlations between adjacent excitation periods limiting the obtained entanglement negativities $N(\rho)$ of Figs. \ref{fig:neg_delay}(b,c) for $(\delta \tau, \delta t) \to 1$. We attribute this effect to the fact that the histogram cross-period correlations are - in contrast to the TTR mode - allowed and cannot be sifted out afterwards. TTR data collection is therefore fundamentally superior to hardware histogramming on the TSPC. Additionally to evaluating cross-correlations also $\XXn$ auto-correlations are shown. These measurements clearly prove the high degree of single photon purity, which equates to an average of $(99.7 \pm 0.2)$ $\%$.\\

\begin{figure}
	\includegraphics[width=1.0\linewidth]{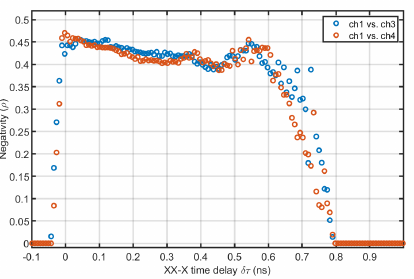}
	
	\caption{Entanglement negativity $N (\rho)$ determined by correlation matrices - c.f. Suppl. Fig. S\ref{fig:P2_corr_matrix} - as function of cross-correlation time delay $\delta\tau$. Detector combinations of channels 1 vs. 3 and 1 vs. 4 are drawn as blue and red circles, respectively.}
	\label{fig:Sup_Neg_vs_Delay}
\end{figure}

Using the cross-correlation data of two different channel combinations of Suppl. Fig. S\ref{fig:P2_corr_matrix} the corresponding entanglement negativities can be evaluated independently for both detector combinations. The result is depicted in Suppl. Fig. S\ref{fig:Sup_Neg_vs_Delay}. Correlations of both channel combinations are recorded simultaneously and share a general outline, still the absolute values vary slightly between the two combinations. This can be understood by considering that the polarization projection on specific bases are not of equal fidelity. Due to the $\Xn$ precession, the contribution of each base to $N(\rho)$ varies as a function of $\delta t$ and therefore also $N(\rho,\delta t)$. This fact also explains why the negativity determined by all four detectors is slightly lower compared to the two detector case, i.e. $N^{(4D)}_\text{max} < N^{(2D)}_\text{max}$. The drop of $N(\rho, \delta\tau)$ beyond $\delta\tau > 0.6$ ns ($> 3 \, T^{\Xn}_1$) can be understood by considering that at these delays only a small fraction of two-photon correlations are recorded and background correlations become increasingly more significant, thereby quenching $N(\rho, \delta\tau > 0.6)$. \\

\subsection*{Blinking analysis}

\begin{figure}
	\includegraphics[width=1.0\linewidth]{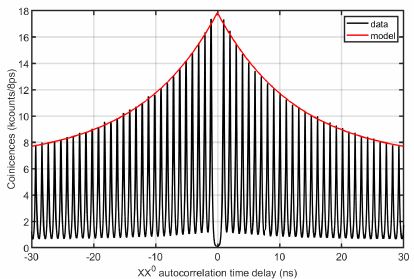}
	
	\caption{Overlay of $\XXn$ auto-correlation trace of HV polarization base as a function of time delay time delay and fitted model adapted from \emph{Jahn et. al.} \cite{Jahn2015}. The fitted model is based on the integrated counts of a excitation period $\TRep$ and is scaled to match the approximate auto-correlation peak heights.}
	\label{fig:Sup_XX0_decay}
\end{figure}

According to \emph{Jahn et. al.} the decay of the auto-correlation function of a QD transition in resonant excitation reveals the typical on-off ratio (i.e. jitter) of this transition \cite{Jahn2015}. The proposed model of this decay is:

\begin{align}
	g^{(2)}_\text{decay}(\tau) = \frac{1-\beta}{\beta} e^{-\frac{\tau}{T_\text{decay}}} g^{(2)}_\text{auto}(\tau) \, ,
\end{align}

where $\beta$ is the on-off ratio, $T_\text{decay}$ is the characteristic decay timescale and $g^{(2)}_\text{auto}(\tau)$ the auto-correlation function without decay. We adopt this model for the case of pulsed resonant excitation by integrating over laser repetition periods $\TRep$. The resulting model of the $\Xn$ auto-correlation data using $g^{(2)}_\text{auto}(\tau) \xrightarrow{|\tau| \geqslant  \TRep} 1$ and $1$ GHz TPE is depicted in Fig. \ref{fig:Sup_XX0_decay} as an overlay. The fitted model yields $\beta = (0.61 \pm 0.02)$ and $T_\text{decay} = (12.7 \pm 0.5)$ ns, which is in strong contrast to the determined excitation efficiency $\ExEff = (0.076 \pm 0.009)$  of the main text. We therefore conclude that the model proposed by \emph{Jahn et. al.} does not accurately describe GaAs quantum dot entangled photon pair sources pumped by TPE. One reason could be that the time dependence of the $\XXn$ transition jitter exhibits a more complicated time dependence (especially for $\delta\tau \to 0$) that is not considered in their model.\\

\bibliography{bibliography}

\end{document}